
\magnification=\magstep1
\baselineskip 18pt
\font\yy=cmbx10 scaled\magstep2
\def\etal{{\sl et al.\/ \rm}}

\def\msun{\,M_{\odot}}
\def\rEI{{\rm E,I}}
\def\rE{{\rm E}}
\def\rI{{\rm I}}

\def\rSNIa{{\rm SNIa}}
\def\rSNIb{{\rm SNIb}}

\def\mmax{m_{\rm max}}

\def\ref{\noindent\hangindent.5in\hangafter=1}

\input tables

{ \nopagenumbers
\null
\vskip 2truecm
\centerline{\yy EVOLUTION OF ELLIPTICAL GALAXIES:}
\medskip
\centerline{\yy I -- THE MULTIPHASE MODEL}
\centerline{\yy }
\centerline{\yy }
\vskip 0.6truein


\centerline{\bf  Federico Ferrini$^1$, Bianca Maria Poggianti$^{2}$}

\vskip 0.3in

\centerline {$^1$Dipartimento di Fisica, Sezione di Astronomia,
   Universit\`a di Pisa}
\centerline {Piazza Torricelli 2, 56100 Pisa, Italia}

\centerline {$^2$Dipartimento di Astronomia, Universit\`a di Padova}

\centerline {Vicolo dell'Osservatorio 5, 35122 Padova, Italia}

\vskip 0.3in
\noindent
\centerline {{\bf Received 27 July 1992}  }
\vfill\eject
}

\pageno=2
\centerline {\bf ABSTRACT}
\vskip 0.5 cm

We present a complex model for the evolution of elliptical galaxies
in the general framework of the multiphase models for galactic evolution.

The elliptical is subdivided into an internal and an external region and
their individual but joined evolution is computed in detail.

Diffuse gas, molecular clouds, stars and remnants are taken into account.
The Star Formation Rate is therefore assumed to be a two--step
process: gas clouds form out of diffuse
gas and then stars form out of gas clouds. Cloud--cloud collisions and
stimulated processes are considered as the main causes
of star formation. The occurrence of winds driven by Supernovae is taken
into account, and the evolution of the system is considered also after
the first wind, allowing for further star formation from the restored gas.

The evolution of the abundances of 15
elements or isotopes (H, D, $\rm ^{3}He$, $\rm ^{4}He$,
$\rm ^{12}C$, $\rm ^{13}C$,
$\rm  ^{14}N$, $\rm ^{16}O$, $\rm ^{20}Ne$, $\rm ^{24}Mg$, $\rm ^{28}Si$,
$\rm ^{32}S$, $\rm ^{40}Ca$, $\rm ^{56}Fe$,
and neutron--rich isotopes synthetized from $\rm ^{12}C$,
$\rm ^{13}C$,  $\rm ^{14}N$ and $\rm ^{16}O$) is followed
with detailed stellar nucleosynthesis as well as stellar
lifetimes are taken into account as a consequence of relaxing the
instantaneous recycling approximation. A new IMF, constant in space
and time, has also been adopted.

The gas removal due to the Supernovae explosions depends on the galactic mass
and the presence of dark matter; the subsequent wind episodes are crucial to
the
intergalactic gas enrichment. Good agreement is obtained for
current SNs rates, Star Formation Rate and gas masses
when compared to the available data.\par

\bigskip\bigskip
{\bf Subject headings:} galaxies: evolution -- stars: formation --
galaxies: stellar content -- nucleosynthesis -- galaxies: abundances

\vfill\eject

\centerline {\bf 1. INTRODUCTION}
\vskip 0.2in

Understanding the evolution of elliptical galaxies from
theoretical arguments involves three fundamental aspects:

(1) formation and organization of the structure throught
a rapid collapse. The dynamical processes during the early evolution of
ellipticals are essential in order to fix the observed properties, more
than for spirals where  one or two -- when considering
the presence of the thin and thick disk components -- accretion phases
determine the large scale structure.

(2) The role of the gas: it is the material from which stars are born; it
is present in a statistically significant sample (Forman \sl et al.\rm, 1985;
Sadler \& Gerhard, 1985; Sparks \sl et al.\rm, 1985; Jura, 1986;
Jura \sl et al.\rm , 1987; Schweizer, 1987;
Bregman \sl et al.\rm, 1988; Huchtmeier, 1988;
Kim \sl et al.\rm, 1988; Sadler, 1988;  Veron-Cetty \& Veron, 1988;
Fabbiano, 1989; Gordon, 1990; Jura, 1990; Di Serego Alighieri
 \sl et al.\rm, 1990;
Walsch \sl et al.\rm, 1990; Gordon, 1991; Roberts \sl et al.\rm, 1991,
Lees \sl et al.\rm, 1991); it is expelled
from galaxies and accumulates in clusters, contributing totally or
partially to the intergalactic gas metal enrichment; it is the source
of the X ray emission; it is a tracer of the gravitational potential;
its origin may be internal or external by capture in the cluster;
a quick substantial removal of gas influences the dynamics inducing a
strong relaxation phase.

(3) Stellar evolution: it gives the amount of  Type I and II supernovae
and then the strenght of the engine responsible for
the heating of the gas and its removal. Nucleosynthesis determines
the enrichment history, the abundances of elements both in the galactic
and intergalactic gas. These quantities depend on Star Formation Rate (SFR)
and Initial Mass Function (IMF).

An evolutionary model may be judged from its ability to joust on various
scales:
the largest for dynamics, the smallest for stars, the intermediate
for star formation and  gas.
In this respect the multiphase approach by Ferrini, Shore and coworkers
gives a good framework,
as it describes satisfactorily the properties of spiral galaxies for
the last two points, with some hints on the first one.
We wish to apply the same physical and mathematical scheme to ellipticals,
considering the characteristic processes of these early type galaxies.

The models previously presented in the literature are not completely
satisfying; in the following we discuss briefly the recent ones.

Arimoto and Yoshii (1987, AY), Yoshii and Arimoto (1987), and Arimoto (1989)
developed a model considering the chemical enrichment  history
as well as the resulting photometric evolution.

Similar models have been further developed by Matteucci and Tornamb\`e
 (1987, MT),
Brocato \etal (1990) and Matteucci (1992) who introduced a more detailed
nucleosynthesis evolution, considering several elements instead of the total
``metal'' content, and a more accurate treatment of the Type I and II SNs,
while
in the AY model only SN II were considered.

Angeletti and Giannone (1990 AG; 1991), studying a model very similar to
the one proposed by AY, criticized AY results on the basis of a non adequate
numerical accuracy in the code.

A common remark to all these approaches is that they do not consider
a self--consistent star formation history: the SFR is assumed simply
proportional to the total amount of gas or to the total gas density,
imposing {\sl a priori} the evolution of
the system. They choose as free  parameter the scaling coefficient
and assume for simplicity a linear dependence.

As it has been already stressed for spiral galaxies (Ferrini \etal 1992, FMPP),
this is a too rough approximation to the problem, since the self--regulating
properties of the star formation (SF) process are now well established
to be not linear and must be considered explicitely in the
physical and mathematical description of the galactic evolution.

Furthermore it is well known that molecular clouds  and not the
diffuse gas phase are the seeds of SF.
On the contrary, the diffuse gas  and not the molecular clouds
suffers efficiently the heating process and consequently can be removed
 from the galaxy,
when the conditions for the occurrence of the wind are established.
Consequently, the two different gaseous phases must be considered
simultaneously
in an evolutionary scheme.

We remember that multiphase descriptions of the interstellar medium
(ISM) have been developed
since many years (see Ferrini 1991a for a review), while only a limited
number of models of galactic evolution considers the presence of molecular
clouds as distinguished from diffuse gas.

All the models in literature, but Arimoto (1989), consider
the SF process ended at the occurrence of the first wind.
The correct treatment of the enriched gas restitution from evolving
stars and the cycling of gas phases as in the multiphase approach
will obviously overcome this rough and unrealistic approximation.

David, Forman \& Jones (1990; 1991a; 1991b) and  Ciotti \etal (1991),
developed hydrodynamic approaches to gas evolution,
where the accent is placed on the heating and expulsion process of the
enriched gas, always with the conventional assumption of an exponential SFR.

Our aim in the present paper is to analyze what can be learnt from the
extension of multiphase models to elliptical galaxies. The following step
will join this multiphase model to a fully hydrodynamic code in connection
with a spectrophotometric code under preparation.

The paper is organized in the following way: in Section 2 we present
the physical ingredients of the model, that is the multiphase scheme
which leads to the set of evolution equations. We discuss in detail the
prescriptions for the Type I and II Supernovae, the explicit conditions
for the wind occurrence and nucleosynthesis prescriptions.
Section 3 is devoted to the presentation of the results for various
computed models. Section 4 deals with the comparison of present
results with the ones obtained from previous models, to underline reciprocal
advantages and caveats. In Section 5 we discuss the solid aspects
of the present study and the improvements we plan to introduce maintainig
the multiphase scheme.

\vfill\eject

\centerline {\bf 2.  THE MODEL}
\vskip 0.2in
This is the first work aimed to model the evolution of elliptical galaxies
with the multiphase approach which gave interesting results for
 spiral galaxies and, in particular, for the solar
neighbourhood (FMPP). The simultaneous study of different phases
of ISM and of chemical evolution, together with the
non-linear mathematical description, allows a self-consistent
calculation of the most meaningful physical quantities.
Such a model allows to introduce easily in the computation
other physical important processes (for example,
heating of gas due to energy released by supernovae).
We follow the physical and computational scenario
used in FMPP; we remind the reader to this paper for details.

\vskip 0.2in
\centerline {\sl 2.1 Interactions of populations}
\vskip 0.2in

We consider two zones inside the galaxy: an ``inner'' region
in the galactic core, whose stars are bound in orbits inside
$1.5$ kpc from the center (Bertin, private communication)
 and an ``outer'' region,
including the whole galactic volume, to which we ascribe all the stars
whose orbits extend out of the internal core.
In this way we partially overcome the absence of an hydrodynamic description
and
may investigate the presence of radial gradients.

The system evolves as a closed box, except for
periods in which a galactic wind occurs and the galaxy loses a part of its gas.
Initial conditions are: homogeneus
sphere of gas of mass $M_G(0)$ and  standard Big Bang chemical composition
(Y=0.24, Z=0.0). The total mass of the system is
$M_{tot}(t)=M_G (t)+M_{out}(t)$
that is the sum of matter still present inside the galaxy
and mass expelled during episodes of galactic wind.
We also explored cases with a dark halo which doesn't partecipate to the
processes of star formation, but contributes to the galactic binding
energy.\par
Different phases of aggregation of matter are considered,
as in Ferrini \& Galli (1988): \par
1. ISM, where we distinguish two components, diffuse gas and clouds.

2. Stars, where we distinguish two sectors: the stars able to induce further
star formation (more massive than $4 \msun$) and the stars
that will not trigger the ISM to condense further into protostars.
Both types contribute in different ways, according to
initial mass, to the chemical enrichment of ISM,
through restitution of matter to the diffuse phase at the end of their life.
The lifetimes and the chemical elements released from
Types Ia, Ib and II supernovae are also included.\par
3. Stellar remnants, the endpoints of star evolution,
that will act as a matter sink, removing mass from chemical evolution.

These populations are kept distinguished for both regions of the galaxy.\par
The evolution of the system is determined by
the interactions between these various  phases; the main physical
 processes we consider
are star formation from clouds (including
cloud--cloud collisions and massive stars--induced
formation); cloud formation from diffuse gas; restitution
of processed material from dying stars to diffuse ISM;
gravitational accumulation of gas from the external galactic
region to the internal one; heating of the gas from
energy released by supernovae; disruption of clouds
for cloud--cloud collisions, for cloud--massive star interactions and
 for supernovae heating.\par
At variance with the other authors,
all these processes are efficient during the whole evolution,
the relative importance of these processes and their self--regulating
 properties determine
the mass fraction of each phase and the chemical enrichment.

\vskip 0.2in
\centerline {\sl 2.2  Supernovae}
\vskip 0.2in
The three observed types of supernovae (Ia, Ib and II) have
been taken into account. They play a fundamental role in the model:
 they contribute to the
chemical enrichment (with products of nucleosynthesis and
times of explosion different from a type to the other) and
they heat the surrounding gas, slowing down the process
of cloud formation and, in case, expelling the diffuse gas from the galaxy.\par
The theorical scenario adopted for the three types is the following:\par
-- {\sl Type II supernovae}: they originate from single massive
stars with mass  $m > 8 \msun$, ending their life
as compact objects (neutron stars or black holes), evolving rapidly
with $\tau(m) < 10^7 {\rm yr}$.

-- {\sl Type Ia supernovae}: they are assumed to originate from
binary systems as described by Iben and Tutukov (1984)
for merging of two degenerate CO stars, after two phases
of common envelope, during which they both lose the hydrogen
envelope. The merging takes place in a time depending
on the loss of energy of the system due to gravitational wave radiation
({\it gwr}) (Landau \& Lifchitz, 1970):
$$\tau_{gwr}=1.48\cdot {{10^8 A^4}\over {M_1\, M_2\, (M_1+M_2)}}$$
\noindent where A is the original separation of the binary system
expressed in $R_{\odot}$,  $M_1, M_2$
the masses of the two components in $\msun$ and $\tau_{gwr}$  in years.
The explosion occurs because in the merging the limit
Chandrasekhar mass is exceeded; $0.6 \msun$ of iron are expelled and no
 remnant is left.\par
The lifetime of such a system is then given by
$$T_{sn}(A\,,\,M)=\tau_{gwr}(A) + T_{ev}(M)$$
where  $T_{ev}$ is the quiescent evolutionary lifetime
of the less massive star of the system; it results as the sum
of the secondary lifetime and of the time necessary for merging, it is then
determined by the two masses and the original separation.\par
-- {\sl Type Ib supernovae}: they are believed to originate from
binary systems, whose initial masses and separation leave
at the end a merging system of a degenerate CO dwarf
and a central He burning star (Iben \sl et al. \rm (1987)).\par
\vskip 0.2in
The difference between the two types of supernovae I
is due in this scenario to the different initial
conditions (masses and separation) of a binary system;
the Ia SNs have a wide range of times of explosion comparable to the Hubble
time
and can then be found also after $10^{11}$  years from the last episode of
star formation. The Ib supernovae, on the contrary, are short-living
systems ( $10^9$ years) and their chemical contribution is rather
different.\par
For these reasons they have to be included separately in the study
of a galactic system such as ellipticals,
which shows large differences in the SFR during the evolution. \par
We computed the Ia and Ib rates as described in
Tornamb\`e (1989). In the total mass range $2-14 \msun$
and initial separation $10 - 10^4 R_{\odot}$, the cumulative number $n$
of exploding systems after a time \sl t \rm from
the episode of star formation is shown in Fig.1,
normalizing to 1 the number of binary systems in the right range of masses
and separations. It has been computed by Tornamb\`e,
according to our choice of IMF and distribution of the
ratio between the primary and the secondary mass, following
the aforesaid theorical prescriptions.\par
It must be stressed that the adopted models for Ia and Ib are two
among various possible binary systems leading to explosion.
Studies of stellar evolution (Tornamb\`e 1989) explored
the theoretical expectations for other types of SNs progenitors, for
which there are not at the moment observational clues.
The model of Ia we adopted has been tested by MT,
giving too low present rates compared to
observations; this could be explained if one assume that
the adopted model is only one of the possible binary
combinations leading to explosion, as suggested by theory.

We compared these results with the rates derived according to Greggio
 \& Renzini (1990)
which assume the model of Whelan \& Iben (1973),
as adopted also by FMPP. Following those prescriptions
 we obtained higher present rates for Ia in ellipticals, in
better agreement with observations.\par
We chose the theoretical scenario given by Tornamb\`e, since it assumes
 physical prescriptions that
seem to be favoured and it is the only one that allows a distinction
 between the two types of SN I.
To test the influence of different SN I rates adopted,
we computed our models using both the rates given by
Tornamb\`e and  introducing, as suggested by Tornamb\`e
(private communication), a parameter able to reproduce the
observed actual rates: such a parameter should take
in account the other possible theoretical \sl channels \rm
of binary systems that could lead to SN esplosion.

\vskip 0.2in
Cox (1972) studied the evolution of a supernova remnant
and determined the evolution of the thermal and kinetic
energy released by each type of SN up the time $t$ after the explosion \par
$$E_{out}=\cases
{0.52 &for $t=t_c$ \cr
1-0.44\cdot {(t_c/t)}^{0.8}\cdot (1-0.41\cdot {(t_c/t)}^{0.41})-
0.22\cdot {(t_c/t)}^{0.8} &for $t>t_c$ \cr }$$
if the initial energy of the remnant is $1$,
being $t_c$ the characteristic cooling time of the shell surrounding
the remnant, taken from us $5.3 \cdot 10^4 $ years.
The total thermal energy of the gas can be computed at each moment as
$$E_{th}(t)=\int_0^t \int_0^{t^*}R(t') \epsilon (t^*-t') \,dt' dt^* $$
Under the assumption that this description is valid
for any type of SN, $R(t')$ is the total rate of SNs at the time $t'$
and $\epsilon(t)$ is the energy per unit of time released
from the SN remnant after a time $t$ from the esplosion. Being  $t^*-t'=\tau$,
$$E_{th}(t)=\int_0^t \Biggl[ \int_0^{t^*} R(t^*-\tau)
 \epsilon (\tau) \, d\tau \Biggr] dt^* $$
We evaluate the integral with a similar procedure to the one described in
FMPP (Appendix).

\vskip 0.2in
\centerline {\sl 2.3  Gas heating and wind}
\vskip 0.2in
The energy released from the SNs to the gas has two effects:\par
i) causes restitution of matter from the phase of cold ISM (clouds)
to the diffuse gas;\par
ii) when the thermal energy exceed the binding
energy of the gas, this can be expelled from the host galaxy.\par
 The first effect can be taken into account introducing
$L$, ratio between the total thermal energy of gas per unit volume and
 density of binding
energy of single cloud, computed from the virial condition:
$\Omega=2 K=2 \cdot 3/2 \cdot n k_B T $
where $K$ and $\Omega$ are the thermal and binding energy of the cloud
per unit  volume, $n$ is the number density of molecules, $k_B$
is the Boltzmann constant and $T$ is the cloud temperature.
For typical cloud values $n=10^3 \rm cm^{-3}$ and $T=20\, K$, we obtain
$$L(t)=E_{th}(t)/1.2 \cdot 10^8 $$
being $E_{th}$ expressed in units of $10^{51}$ ergs.

The second effect is switched on at the moment the condition
$$ E_{th}\,\ge \,E_B \eqno(1)$$
is satisfied, i.e. when the thermal energy of gas exceeds its binding energy.
This one depends on the total galactic mass and on the gas mass (Saito, 1979):
$$E_{B_{gas}}=16.43 \cdot 10^8 {(M_G)}^{1.45}\cdot {{M_{gas}}\over {M_G}}
    \Biggl(2-{{M_{gas}\over {M_G}}}\Biggr) $$
$E_{B_{gas}}$ is expressed in units of  $10^{51}$ ergs;
the galactic mass  $M_G$ and the gas mass $M_{gas}$ are in
units of  $10^{12} \, \msun$. In presence of dark matter,
$M_G$ represents the sum of the luminous and dark matter, if this is
distributed
like the luminous one.\par
The gas is assumed to be lost from the galaxy at a decreasing
exponential rate and it will continue to be expelled
until the condition (1) is satisfied, but at any moment the restoration
of diffuse gas from evolving stars is still computed.\par
When the mass of the restored gas
makes the binding energy of the diffuse ISM higher than
the thermal energy, the wind interrupts and a phase of ``quiescent''
 evolution follows.\par
Depending on the rates of SNs and the amount of gas released
from evolving stars, other episodes of wind can occur.\par

\vskip 0.2in
\centerline {\sl 2.4  Nucleosynthesis prescriptions}
\vskip 0.2in
Abundance evolution of 15 elements or isotopes is computed:
H, D, $\rm ^{3}He$, $\rm ^{4}He$, $\rm ^{12}C$, $\rm ^{13}C$,
$\rm ^{16}O$, $\rm ^{14}N$, $\rm ^{20}Ne$, $\rm ^{24}Mg$,
$\rm ^{28}Si$, $\rm ^{32}S$, $\rm ^{40}Ca$ , $\rm ^{56}Fe$
and neutron-rich isotopes synthesized from $\rm ^{12}C$,
$\rm ^{13}C$, $\rm ^{14}N$ and $\rm ^{16}O$.\par
Starting from an initial composition of $24 \%$ $\rm ^{4}He$
and the remaining hydrogen, the evolution of each element is
followed using the \sl production matrices \rm $Q_{ij}(m)$, first
introduced by Talbot \& Arnett (1973), that give
the fraction of the mass of an element $j$ initially
present in a star of mass $m$ that is transformed in element $i$ and ejected.
A detailed description of the way these matrices are
computed and of the nucleosynthesis prescriptions is given in FMPP.\par
With respect to their elements production,
the only difference is in the nucleosynthesis
of SN I: we assume for SN Ia the prescriptions given
by Nomoto \sl et al. \rm (1984) and for SN Ib by Branch \& Nomoto (1986).
The Type Ia SN's are in this way assumed to produce $0.6 \msun$
of iron, together with smaller quantities of
other heavy elements (C, O, Ne, Mg, Si, Ca), and the Ib's expel $0.3 \msun$
of iron and the rest as $\rm ^{4}He$; both supernovae
are assumed to leave no remnant.\par

\centerline {\sl 2.5  The equations}
\vskip 0.2in
According to the different phases described in section 2.1, we define \par
$$\vbox {\halign{ $\hfil#$&&${}#\hfil$&\qquad$\hfil#$  \cr
           g &=\displaystyle{ M_{\rm gas} \over M_{\rm tot}},
          &c &=\displaystyle{  M_{\rm clouds} \over M_{\rm tot} } \cr
   \noalign {\medskip}
           s &=\displaystyle{  M_{\rm stars} \over M_{\rm tot}},
        &r &=\displaystyle{  M_{\rm remnants} \over M_{\rm tot}} \cr}}$$
The system of equations we consider is:\par
$$\eqalign {{ds_{1\rE} \over dt} &= H_1 c_\rE^2 +a_1 c_\rE s_{2\rE} -
 D_{1\rE} \cr
{ds_{2\rE} \over dt} &= H_2 c_\rE^2 +a_2 c_\rE s_{2\rE} - D_{2\rE}  \cr
{dg_\rE \over dt} &= -\mu (1-L) g_\rE^n + a'c_\rE s_{2\rE} + H'c_\rE^2-
 fg_\rE + W_\rE  \cr
{dc_\rE \over dt} &= \mu (1-L) g_\rE^n - (a_1 + a_2 + a')\,c_\rE s_{2\rE}
-(H_1 + H_2 + H')\,c_\rE^2 \cr
{ds_{1\rI} \over dt} &= H_1 c_\rI^2 + a_1 c_\rI s_{2\rI} - D_{1\rI} \cr
{ds_{2\rI} \over dt} &= H_2 c_\rI^2 + a_2 c_\rI s_{2\rI} - D_{2\rI} \cr
{dg_\rI \over dt} &= -\mu (1-L) g_\rI^n  + a'c_\rI s_{2\rI} + H'c_\rI^2 +
fg_\rE
      + W_\rI \cr
{dc_\rI \over dt} &= \mu (1-L) g_\rI^n  - (a_1 + a_2 + a')\,c_\rI s_{2\rI}
- (H_1 +H_2 + H')\,c_\rI^2  \cr
{dr_\rE \over dt} &= D_{1\rE}+ D_{2\rE} - W_\rE \cr
{dr_\rI \over dt} &= D_{1\rI}+ D_{2\rI} - W_\rI \cr
{d X_{i\rE} \over dt} &= ( W_{i\rE} - X_{i\rE} W_\rE) /( g_\rE + c_\rE ) \cr
{d X_{i\rI} \over dt} &=
(W_{i\rI} - X_{i\rI}W_\rI + fg_\rE(X_{i\rE}- X_{i\rI}))/
      (g_\rI + c_\rI) \cr } $$
\bigskip
The subscripts E, I denote external and internal quantities. In the $s$ phase
we distinguish two classes, corresponding to different stellar mass ranges:

$$\eqalign{ s_1 &: \cr s_2 &: \cr } \quad
  \eqalign{ m_{\rm min}< &m \le m^*    \cr
            m^* < &m \le m_{\rm max} \cr }$$
because only $s_2$ {\sevenrm (massive)} stars interact with clouds inducing
star formation; $m_{\rm min},\, m^*,\,$ $m_{\rm max}$ are estimated to be
$0.1\msun$,\ $4\msun$,\ $100\msun $ respectively.
\smallskip

$X_i$ are the abundances by mass of the fifteen chemical elements.

The processes and the related coefficients considered are:\par
1. Star formation from cloud--cloud collisions: $H_{1,2}\, c_{\rE, \rI}^2$ \par
2. Induced star formation, due to the interactions
between massive stars and clouds: $a_{1,2}\, c_{\rE, \rI} s_{2\rE, \rI}$ \par
3. Cloud formation from the diffuse gas (inhibited from
the supernovae driven gas heating) : $\mu (1-L) g_{\rE, \rI}^n $ \par
4. Diffuse gas restitution from cloud--cloud collisions: $H'\, c_{\rE,
 \rI}^2$ \par
5. Diffuse gas restitution from induced star formation:
$a' \, c_{\rE, \rI} s_{2 \rE, \rI}$ \par
6. Enriched diffuse gas restitution from evolving stars: $W_{\rE, \rI}$ \par
7. Infall of matter from external to internal zone
from gravitational accumulation: $f g_{\rE}$ \par
The determination of the coefficient $L$ has already
been described in Sect. 2.3; the rate coefficients expressing
the microphysics of the process  have been evaluated as the ratio of
 the efficiency
to the typical timescale of the physical process considered,
as described in FMPP. In the case of ellipticals,
it has been considered that velocity dispersions are higher by
factor 10 than the ones observed in the solar neighbourhood,
and this reflects on the typical timescale of clouds collisions
($H_{1,2}$ ten times higher than in the case of spiral galaxies).
The coefficients are in this way evaluated, in a similar way
to what has been done for spirals, from the observations
and from the detailed theory of star formation and evolution.
The parameter $f$, connected with the accumulation of gas towards
the center, (the only true free parameter in the case of spirals),
has been estimated considering the higher galactic masses
of ellipticals and requiring that the observed mass ratio
between the external and the internal regions has to be reproduced.\par
The Star Formation Rate determined self-consistently in any phase of
 the evolution is:
$$ \eqalign {\psi_\rE (t) &= (H_1 + H_2)\, c_\rE^2 + (a_1 + a_2)\,c_\rE
 s_{2\rE} \cr
\psi_\rI (t) &= (H_1 + H_2)\, c_\rI^2 + (a_1 + a_2 )\, c_\rI s_{2\rI} \cr}
$$

Here the reader can see one of the fundamental differences among our model
and AY, MT, AG models: the SFR is the principal consequence of the evolution
of the system and not an ``a priori'' choice. All these authors consider
the SFR linearly proportional to the amount of total (AY, AG) or surface
density (MT) gas. They assume the same expression for the efficiency of SF:
$\nu \propto M_G(0)^{-0.115}$, so an efficiency which decreases with the
total galactic mass, at variance with the results of Tinsley \& Larson (1979)
which explained the mass-metallicity relation as due to the increasing
 efficiency
of SF with total galactic mass.

The system of coupled non--linear differential equations
is valid when the condition for a wind
doesn't occur. Any time equation (1)
is verified, a term is added to the equation of the gas phases
in both zones, introducing a loss of gas from the galaxy at a
decreasing exponential rate $\tau_i = 1/k_i$:
$$\eqalign
{{dg_\rE \over dt} &= -\mu (1-L) g_\rE^n + a'c_\rE s_{2\rE} +
 H'c_\rE^2- fg_\rE + W_\rE  - k_1 g_\rE \cr
{dg_\rI \over dt} &= -\mu (1-L) g_\rI^n  + a'c_\rI s_{2\rI} +
 H'c_\rI^2 + fg_\rE+ W_\rI  - k_2 g_\rI \cr}$$

The contribution of each star depends on the initial mass and on the
 characteristics of
the binary system to which it belongs, in the case of SNs I.
The Initial Mass Function (IMF) we assume is based on the studies of
Ferrini, Palla \& Penco (1990) on the fragmentation of molecular
clouds in the solar neighborhood and the adopted expression (Ferrini, 1991b)
is considered valid in any zone of the galaxy and at any time:
      $$ \varphi (m) = 2.01\,m^{-.52}\, 10^{-[2.07 (\log m)^2 +
        1.92 \log m + 0.73]^{1/2}}  $$
For a detailed description of the mass distribution in the binary
systems we refer to FMPP.

The death rates $D_{1\rE}$, $D_{2\rE}$, $D_{1\rI}$, $D_{2\rI}$
represent the fraction of mass that leaves the star phase
at the time $t$ in the two zones and for low-mass and massive
stars:\par
$$ D_{1E,I}(t) =\int \limits_{m(t)}^{m^*} \psi_{E,I}
 (t-\tau_m)\varphi (m) dm  $$

$$ D_{2E,I}(t) =\int \limits _{m^*}^{m_{max}}\psi_{E,I}
(t-\tau_m)\varphi (m) dm  $$

The mass $m_{max}$ is the maximum stellar mass
and it is taken $100 \msun$;  $m(t)$ is the star mass
for which the lifetime $\tau (m)$ is equal to $t$.\par
The restitution rates $W_{i E,I}(t)$ give the fraction of mass expelled
 into the ISM
at the time $t$ as element $i$ (processed or unprocessed),
in the two zones. They are the fundamental terms to compute
the gas mass and chemical composition. They can be written as \par
$$ W_{i;\rEI}(t)=
\int \limits _{m_{\rm min}}^{m_{\rm max}}
\left[\sum_j \tilde Q_{ij}(m) X_{j;\rEI}
(t-\tau_m)\right]
 \psi_{\rEI} (t-\tau_m)\,dm \eqno (2)$$
The term $\tilde Q_{ij}$ includes the contribution
from quiescently evolving stars, SNs II, Ia and Ib
 and can be explicitely introduced in eq. (2)

$$ \eqalign {
 W_{i;\rEI}(t) =  \int\limits_{m(t)}^{\mmax} \left\{ \vphantom{\sum_j} \right.
    &  \left. \left[ \sum_j Q_{ij}(m) \ X_{j;\rEI}(t-\tau_m) \right]
       \psi_{\rEI}(t-\tau_m)\, \varphi (m)\,+ \right. \cr
   + &  \left. \left[ \sum_j Q^{\rSNIa}_{ij} \ X_{j;\rEI}(t-\tau_m) \right]
  \psi_{\rEI}(t-\tau_m)\,\cdot 1.4\cdot N_a (m) + \right. \cr
   + &  \left. \left[ \sum_j Q^{\rSNIb}_{ij} \ X_{j;\rEI}(t-\tau_m) \right]
  \psi_{\rEI}(t-\tau_m)\,\cdot 1.4 \cdot N_b (m)  \right\} \, dm  \cr } $$
The first term on the right side shows the contribute
of the quiescent evolving stars and of SNs II; the second and the third
term indicate respectively the SNs Ia and Ib contributes: the matrices
 $Q^{\rSNIa,\rSNIb}$ give the restituted elements from the two types
of SNs. These matrices do not depend on the star mass,
being in any case $1.4 \msun$ the mass involved.
$N_{a,b}(m)$ represents the number of SNa events occurring when $1 \msun$ of
gas is transformed into stars at the time $\tau (m)$.
This is $N_{a,b}(m)= \eta n'(\tau_m)$ where  $n'(t)={{dn}\over{dt}}$,
number of SNs exploding after a time $t$ after the formation of the
 progenitor system.
The delay between the star formation and the release
of chemical elements to the ISM is taken into account for
quiescent and exploding stellar systems; for normally
evolving stars and for SNs II this time is taken equal
to the main sequence permanence time $\tau (m)$, for
which we assume (Burkert \& Hensler, 1987):\par
$$ \tau_m =\cases {8.0\times m^{-2.8}\ \ \ {\rm Gyr}\quad
      &for $\quad m\leq 10.0\msun$ \cr
                  0.05\times m^{-0.6}\ \ {\rm Gyr}
      &for $\quad m > 10.0\msun $.\cr }$$
Finally, the supernovae rates are evaluated as follows:\par
$$R_{II}(t)= \int_{8 \, \msun}^{m_{max}} { {\varphi ' (m)} \over m}
 \psi_{\rE, \rI} (t-\tau_{m}) \, dm $$
and
$$R_{Ia, Ib} (t)=\eta \int_0^t \psi_{\rEI} (t') {n'_{Ia, Ib} (t-t')} dt' $$
where $\varphi ' (m)$ is the IMF including both single stars
and primary massive stars belonging to bynary systems that give
rise to SN I events (see FMPP Appendix).\par
In the rate of SNIa,b the coefficients $\eta$ represents the
number of bynary systems in the range of total mass
$2< M < 14 \, \msun$ and initial separation $10< A < 10^{4} R_{\odot}$
for  $1 \, \msun$ becoming star. This has been deduced from observations,
 considering that
about half of the formed stars are binary systems and that the observed
 separation
distribution of binary systems is $A \, \propto \log A$  (Popova \sl et al.
 \rm 1982).\par

\vskip 0.2in
\centerline {\bf 3. RESULTS}
\vskip 0.2in

Observations concerning elliptical galaxies are not as rich as the ones
concerning the solar neighbourhood, for which we have a detailed
knowledge of stellar populations, of gaseous phases, of
abundances both in stars and in gas. Our effort to model the Milky Way
was devoted mainly to the local region, where many constraints must be
considered. Elliptical galaxies are in this respect a completely different
framework: we can appreciate here the fact that multiphase models, starting
from the knowledge of physical interactions on a limited scale, allow to infer
large scale properties and global trends as observed in ellipticals;
this is a non--trivial result, at variance with previous models who assumed
in principle the large scale relations between the interesting quantities.
The self--regulating nature of galactic evolution well considered in the
multiphase approach is the support of our physical and mathematical structure.

In Table 1 we show the results of some selected models; model A
is taken as the standard reference model and corresponds to
a galactic system of mass $M_G (0) = 10^{12} \msun$
without dark matter and with a high efficiency of gas loss
($k=1000$ corrisponding to a timescale $\tau =10^4$ years).
Models B and C show the results in the case of less important
episodes of galactic winds (due to a lower gas loss
efficiency $k$ and to a lower SNs Ia rates respectively).
The last three models refer to galactic systems with
different initial mass or in presence of dark matter.
Masses are in units of
$\msun$ while the third column shows the
coefficient $k$, inverse of the gas espulsion timescale
$\tau$ expressed in $10^7$ years.\par
The forth column of Tab.1 lists the time of the first galactic wind
in Gyr, in the fifth column $M_{out}$ is the mass fraction
lost for all the occurring winds, the sixth column shows the amount of iron
expelled during the whole evolution in $10^8 \msun$, following columns
report the  present SFR ($\msun  \, {\rm yr}^{-1}$); the time, in Gyr, of the
last wind; the present rate (SNU) of SN Ia and the mass fraction
related to the initial total mass at the end of evolution (after 16 Gyr)
in the phase of gas, clouds, living stars and remnants in the whole galaxy.\par

\vskip 0.2in
\centerline {\sl 3.1 Star formation and galactic winds}
\vskip 0.2in
The Star Formation Rate during the first 2  Gyr for the standard model
A is shown in Fig.2. The maximum SFR occurs $10^8$ years after
galactic formation and amounts to $1600 \msun {\rm yr}^{-1}$, in agreement
with the maximum SFR invoked by Sandage (1986)
to explain the different morphological types along the Hubble sequence. \par
The SFR remains higher than $1 \msun {\rm yr}^{-1}$ for the first 2.5 Gyr.

As shown in Tab.1, the present SFR strongly depends on the efficiency of
galactic winds: the too low SNs Ia present rate
obtained using original Tornamb\`e calculations (model C) gives a quite
high SFR compared to the observed one
($0.1 - 1  \, \msun {\rm yr}^{-1}$ , Thronson \& Bally, 1987).
The comparison with the observed SFR in normal ellipticals
put a constraint also on the efficiency
of the gas loss: values of $k$ giving mass loss timescales
shorter than $10^4 - 10^5$ years give a too high  SFR
compared to  observations (model B).\par
While the ratio between internal and external SFR during the
first phases of the evolution is $10^{-5}$, the two present
values are comparable and this fact is due to the gas
accumulation towards the center.\par
The $90 \, \%$ of star formation takes place during the first
Gyr, but the evolution of these systems is much different
from a passively evolving galaxy: the amount of gas
restored from evolving stars and the later consequent SFR
depend in fact on the following episodes of wind.
The present SFR in solar masses per year increases with the final
galactic mass (in units of $10^9 \, \msun$) as:
$$SFR=0.073 \cdot {(\log M_f)}^2 -0.129 \cdot \log M_f + 0.067$$
In the case of the model A, the galaxy remains under wind
conditions between 0.35 and 1.36 Gyr and later winds also
occur. While the first episodes are driven by SN II, the
smaller but still not null amount of SN Ia later explosions
is responsible for subsequent gas loss.\par
Less massive galaxies are subjected to sooner and more
majestic mass losses through winds: from Tab.1 is evident
that this does not  correspond to a higher iron enrichment
of the intragalactic cluster medium (IGM) from less massive galaxies.
Massive systems expell their gas at later times and their ISM had, in this way,
time to be enriched from SN I explosions, whose timescales
are in average longer than the SN II ones, responsible for the wind.

The loss of heavy elements from ellipticals
can give a contribution to the  IGM, where high
quantities of iron have been observed (Vigroux, 1977; de Young, 1978;
White, 1991; Arnaud \sl et al \rm, 1992 and references therein):
there is, in fact, evidence that the iron mass in the IGM is directly
correlated with the stellar mass in the early type
galaxies of the cluster, as well as the total IGM gas mass.\par

It is possible to recognize a regular behaviour of the winds with
the initial galactic mass. Higher is the mass, later comes the first wind,
lower is the fraction of expelled mass, higher is the amount of heavy elements
in the removed gas. At the end of evolution, the smaller galaxies are always
in a wind phase, while the more massive ones are in a regime
 of accretion towards
the center, even thought the SN I rates are higher for more massive galaxies.
Naturally the total number of winds is anti--correlated to the initial
 galactic mass.\par
The amount of iron and oxygen expelled in total increases with the final
galactic mass as:
$$\log M_{Fe}=-0.51 \cdot {(\log M_f)}^2 +2.97 \cdot \log M_f -4.08$$
where both are expressed in units of $10^9 \, \msun $.
The same relation for the oxygen is:
$$\log M_O=-0.56 \cdot {(\log M_f)}^2 +3.19 \cdot \log M_f -4.24$$.

\vskip 0.2in
\centerline {\sl 3.2 Supernovae}
\vskip 0.2in

The frequency of supernovae explosions has been found
from observations for different morphological galactic types;
Cappellaro \& Turatto (1988) derived a rate of 0.09 SNU
(number of supernovae per $10^{10} \, L_\odot$
blue galactic luminosities per 100 years)
for a sample of ellipticals from the Second Reference Catalogue
of Bright Galaxies (de Vaucouleurs \sl et al.\rm, 1976),
adopting a value of $ H_0=50$ km s$^{-1}$ Mpc$^{-1}$.
A more recent analysis of Turatto e Cappellaro (private
communication), found a rate of 0.07 SNU for ellipticals
from the Nearby Galaxies Catalog (Tully, 1988)
and of 0.1 SNU for the Third Reference Catalogue of Bright
Galaxies  (de Vaucouleurs \sl et al.\rm, 1991),
 with a value of $H_0=75$ km
s$^{-1}$ Mpc$^{-1}$ (new results are lower by a factor 3
from the old ones).\par
Different rates are found by Evans \sl et al.\rm (1989):
$0.32 \, h^2=0.18 $ SNU if $h=75$
in a sample of early type galaxies that doesn't
distinguish between E and S0. Adopting a ratio between the frequencies
in E and in S0 as Cappellaro \& Turatto (1988) found,
this corresponds to 0.05 SNU for ellipticals.\par
Van den Bergh \& Tammann (1991), applying some corrections to Evans
rates, found for E+S0 the value $0.98 \, h^2$ SNU, that for $h =75$
and the ratio between E and S0 rates already mentioned, gives 0.15 SNU
 for ellipticals.\par
The low number of supernovae observed (3 in ellipticals
for RC3 and 1 for the Tully Catalogue) and the strong dependence
on the assumptions and on the Catalogue
suggest that these rates are to be taken only as indicative,
before an extensive systematic survey is completed.
Moreover, the classification of early-type galaxies can
often be ambiguous and this could affect the estimated rates.

The computed present rate of SNIa of column 9 in Tab.1  for
most of the models is in good agreement with the observational
estimates. To convert from internal units to SNU we have
used a value of $M/L=8$, as derived from dynamics studies (David,
Forman \& Jones, 1991).

The rates of SN II and Ib follow the SFR: they reach the maximum
value just before the onset of the first wind (both reaching
$2.5$ SNU) and drop fast to low
values until they become negligible after 16 Gyr compared to the SN Ia
present rate. The  SNIa reach their maximum almost at the same time
the other two types do, but decrease more slower and remain the only
SNs observable at present time.\par
{}From Tab.1, the model with higher SNIa present rate is model B,
the one with reduced gas loss efficiency (low $k$), because of the higher
SFR in later times; model C, adopting rates according
to Tornamb\`e without correction to reproduce the same rates
given from the model by Greggio \& Renzini, shows too few SNIa's at
 the present time.\par
The amount of Type Ia SNs (in SNU) at the present time increases with the final
galactic mass (in solar masses) according to the relation:
$$R_{SNIa}=0.006 \cdot {(\log M_f)}^2 -0.008 \cdot \log M_f +0.030$$

\vskip 0.2in
\centerline {\sl 3.3 Gas, clouds, stars and remnants: galactic morphology}
\vskip 0.2in
The mass fraction in the different phases (gas, clouds, stars
and remnants) is shown in Fig.3  and Fig.4
for the two regions.
The present mass fractions are presented in Tab.1;
the amount of warm diffuse matter found with our
models is in good agreement with the recent observations
of $10^8 - 10^{11} \msun$
of hot X-ray emitting gas (Ciotti \sl et al. \rm 1991 and references therein),
that appear a common phenomena in luminous ellipticals.
Recent detection of HI and first measurements of molecular
species show that a not negligible fraction of ellipticals
have  cold ISM in a wide range of masses (Schweizer, 1987).\par

In some cases there is evidence that the cold ISM is the result of
capture (Sage \& Galletta, 1992),
while the internal component seems to be the hot X-ray
gas, whose luminosity often correlates with general galactic
properties. Anyway, the sometimes ambiguous classification
of early type galaxies could be misleading in understanding
the real cold component of the ISM in ellipticals.\par
The amount of present total diffuse matter (gas + clouds) increases with
the final galactic mass according to the relation:
$$\log M_{g+c}= 0.80 \cdot \log M_f - 1.40$$
where the masses are expressed in $10^9 \, \msun$.

This trend is confirmed from observations:
as shown by Fabbiano \sl et al. \rm (1992) in their catalog
of galaxies observed in the soft X-ray band with the
\sl Einstein Observatory \rm, the X-ray emission of
early type galaxies increases with the total \sl B \rm
luminosity: $\log L_X= 1.8 \log L_B
+21.5$, where $L_X$ is expressed in $\rm erg \ s^{-1}$ and $L_B$
in solar luminosities.
{}From these data is possible to derive the X-ray gas mass according to
the relation given by Canizares \sl et al. \rm (1987):
 $\log M_X= 0.5 \log L_X (watts) +1.2 \log (L_B/L_{\odot}) -20.5$
($M_X$in solar masses)  and transforming
into our units $(M/L_B =10)$ we obtain the relation between hot gas
and galactic masses as derived from observations (in units of $10^9 M_{\odot}$:
$$ \log M_X= 2.1 \log M_{gal} -5.45.$$
The linear correlation we found agrees with the trend
deduced from observations, while the sample of early
galaxies in which cold ISM is observed
is still too small to allow a quantitative investigation
for the total amount of diffuse matter (Roberts \etal 1991).

\vskip 0.2in
\centerline {\sl 3.4 Chemical evolution}
\vskip 0.2in
Detailed chemical evolution of the 15 species studied
has been computed: the enrichment in ellipticals appears
to be a faster process than in spirals, due to the high
efficiency of star formation of the early phases.
Not only these systems seem to be more efficient in
producing heavy elements, they also do it with relative
abundances different from our galaxy and,
in general, expected spirals. This is evident from the
comparison between Fig.5   and Fig.6 taken from FMPP.
This effect is determined from the higher ratio between
SN II (producing mostly O and Mg) and SN I (Fe productors)
during the early evolution of ellipticals,
while the opposite happens after the first wind.\par
We have also calculated the total metal content Z,  sum
of all elements heavier than $\rm ^{4}He$: the result
is shown for our standard model in Fig.7.

The monotonic increasing metal content with age found for the Milky Way
(FMPP and Fig.8), does not appear in
ellipticals: after the onset of the first galactic winds,
typically around 1 Gyr from galactic formation, the metal
content increases for the restitution of intermediate massive
stars and decreases in the following evolution
for the dilution from chemical restitution of less massive stars.\par
This phenomenon is not present in spirals, where the restituted
matter is diluted from the primordial metal-poor diffuse matter,
present in high quantities,  not swept--away from winds and
not too much consumed from star formation.\par
Quantitative chemical comparison with observations are
quite dearing to do.
Most of the observations about chemical elements in ellipticals
are found between the index $Mg_2$ and other fundamental characteristics,
such as luminosity, central velocity dispersions ecc.
Although these relations are frequently studied, they cannot be easily
understood in terms of a chemical evolutionary hystory.\par
The calibration of $Mg_2$ against [Fe/H] normally used
(Burstein, 1979), based on globular clusters observations
and stellar models with solar relative abundances, is expected to
fail in reproducing the relation existing in ellipticals.
Furthermore, the observed spectral indices are produced from
the contribution of all stars, averaged with their luminosities.\par
Anyway, two observed trends can be singled out, even though
it is difficoult to transform them in chemical abundances of single
star generations: \par
a) the colour-magnitude relation: the average galactic metal content
increases with the total galactic mass.
Yet, recent observations have found high mass ellipticals
with extraordinarily low metal index $Mg_2$ (Bertola \sl et al. \rm, 1992).\par
b) the radial metal gradient: as in spirals, more internal regions
appear more metal rich (Pagel \& Edmunds, 1981).\par
A following paper, presenting the spectrophotometric evolution
computed according to our model, will analyse these two aspects.
\vfill\eject
\vskip 0.2in
\centerline {\bf 4. COMPARISON WITH PREVIOUS MODELS}
\vskip 0.2in
In order to analyse some fundamental differences
between ellipticals and spiral galaxies,
we compare our results with the standard model of FMPP
for the solar neighbourhood. Both models use the same
multiphase approach and a detailed chemical description.
\par
Our results will be compared also with the models
of Matteucci \& Tornamb\`e (MT), Arimoto \& Yoshi (AY)
and David, Forman \& Jones (DFJ).
\vskip 0.2in
\centerline {\sl  4.1  Model for spiral galaxies }
\vskip 0.2in

The star formation and, consequently, chemical histories
of spirals and ellipticals are definitely different.
The SFR for our Galaxy as computed by FMPP in their
standard model shows a maximum SFR of $\sim 8 \, \msun$
both for halo and galactic disk, to be compared with the SFR
in Fig.2 of $\sim 1600 \, \msun$ and $\sim 23 \, \msun$
respectively for external and internal regions.
In both cases, the accreting regions (internal for
ellipticals and disk for spirals) show a maximum delayed with
respect to the outer zone; the SFR is kept
non-null from the accretion of gas, maintaining a value of
$1 \, \msun$ in the present galactic disk and of about
a tenth at the center of a massive elliptical.\par
Fig.7 and 8  show the total metal content in the ISM
(percentual of mass in elements heavier than $^4$He)
for our Galaxy and for model A.
Due to the high SFR in ellipticals, the enrichment of
ISM from restitution of matter from evolving star
has a timescale much shorter than in
solar neighbourhood: in model A a solar metallicity
(corresponding to 10 Gyr in the spiral model) is
reached in the external region within 1 Gyr
from galaxy formation. The metallicity of an elliptical
is \sl besides this very short first evolutionary
phase, always super-solar; \rm moreover, both for the Milky
Way and ellipticals, the accreting region (disk or
internal) has higher present metallicities, due to the
gravitational deposit towards the center of gas
enriched from stars evolving in the outer zone;
while in the solar neighbourhood the metallicity of the disk
becomes higher than in the halo soon after the galactic formation
(within half Gyr), the outer zone in ellipticals remains
more metal-rich during a longer period (6 Gyr for model A).

Ellipticals are systems where the chemical enrichment
is accelerated with respect to spirals: considering the chemical different
histories, it appears difficult to explain the high metallicities
observed in ellipticals if they originate
from merging of spirals, at whichever phase of their evolution. The high
initial SFR that can explain the morphological appearance
of these systems is able, instead, to reproduce such metallicities.\par
Fig.7 shows a peculiar phenomenon due to the existence in ellipticals
of galactic winds: while in our
galaxy observations and models predict a monotonic growth of
the metallicity with time, after the onset of the
first episodes of galactic
winds, Z increases quickly to very high values (maximum 2.5
times solar), because of the evolution of massive stars
whose processed elements do not mix with  a primordial metal-poor
ISM. In the following evolution,
less massive stars evolve and release matter to the diffuse phase,
and the metal content Z decreases.\par
Another difference between the chemical enrichment in our Galaxy
and in ellipticals is evident from Fig.9.
In both cases, oxygen is predominant on iron
during the early phases, because of the
initial SNs II rate. The following evolution is faster in
ellipticals also in this case: the iron produced by SNs I
and the absence of SNs II (oxygen producers) after the first Gyr
produce the situation showed in Fig.5: the ratio [Fe/H]
reaches values higher than solar before the end of the
first Gyr, while the great quantity of oxygen
produced by the numerous initial SNs II has been expelled
with the winds and the ratio [O/H] in the ISM attains
much lower values than in our Galaxy (-1.2 dex in the
external region against the present -0.1 dex of the solar
neighbourhood).\par
This different contribution of iron and oxygen in systems of
various Hubble type is evident  in Fig.9 (a,b), where
[O/Fe] versus [Fe/H] is plotted for the external region
of model A and the galactic disk as computed by FMPP.
The situation in the internal region takes more time to settle because of
the enrichment from the external zone.\par
Finally, also the trend of the helium content with increasing
metallicity is different in spirals from ellipticals:
for the solar neighbourhood, FMPP found a monotonic growht
of $^4$He with Z and in all the literature the helium enrichment
parameter ${\delta Y}\over {\delta Z}$ is considered to be zero or
a positive constant value. On the contrary, in ellipticals,
the trend of helium versus Z is not monotonic, because of
the previously described chemical behaviour.
This second phenomenon must be taken
into account in population and spectroscopic studies, because
the ratio Y/Z can change drastically the stellar properties
considered (\sl i.e. \rm the latest stages of the evolution
of low mass stars, fundamental to understanding the progenitors
of the UV rising branch observed in ellipticals, change, for
a given age and metallicity, with the precise trend
of the helium content with metallicity (Greggio \& Renzini, 1990)).

\vskip 0.2in
\centerline {\sl 4.2  Previous models of elliptical galaxies}
\vskip 0.2in

The same general criticism presented in FMPP with respect to other
spiral galaxies evolutionary models can be extended to elliptical
models. A major difference between our approach and  previous models of
ellipticals is the self--consistent treatment of star formation process.
This is a unique feature of the multiphase model, where we are not forced to
choose a priori a SF history, which obviously forces the I and II SNs
 production,
irrispectively of the evolution of the system. This approach allows
 us to consider
the star formation, although drastically reduced, at times successive to
 the early
winds; this indeed will not necessarily curtail the SF, as indicated by the
presence of starburst galaxies with X--ray winds and ongoing massive star
formation (see DFJ for references).

The necessity of a self--consistent modelling of the development of
 galactic winds
and continued star formation is underlined in DFJ. Only Arimoto (1989) has
discussed models for which the SF has not been interrupted after the
 first wind, but his
treatment is unsatisfactory for the simplicistic star formation "law" and for
the absence of a cloud phase.

Our model, so, clashes with AY and MT models, whose results cannot
 in principle be
compared with the observations of ellipticals, while is complementary to the
DFJ studies.

Notwithstanding the differences in the methods, a few qualitative
 behaviours result the same
in the various models: the increase of first wind occurrence time and
 the decrease of expelled
mass fraction with galactic mass, the fraction of remaining gas, the
 influence of dark
matter which increases the time of the first wind. Like AY we find
 a metal distribution
in stars with similar extremes and as MT we get different correlations between
the single elements expelled and galactic mass.

Naturally, differences are present, for example our $t_{gw}$ are shorter than
the AY and MT ones, while our expelled masses are larger.

The relation between the time of the first wind $t_{gw}$,
expressed in Gyr, and the initial galactic mass $M_G(0)$
in solar masses is: $\log t_{gw}=0.37 \cdot \log M_G(0) -4.89$.
The relation found by MT was: $\log t_{gw}=0.35 \cdot \log M_G(0) -4.41$
for Salpeter IMF, while AY found $\log t_{gw}=0.44 \cdot \log M_G(0) -5.42$.

The continued presence of winds at later times generates also differences
in the elements pollution into the intergalactic gas, which increases, although
slowly, at late times. The absolute amount of material is low but its
 metal content
is high and its contribution to the extragalactic abundances of heavy
 elements is
relevant, in agreement with DFJ results.

An important qualitative comparison concerns the regimes of gas
 expulsion, as function of
galactic mass. We are in agreement with DFJ results for what concerns both
low mass galaxies, able to maintain galactic winds for most of their lifetime,
and massive galaxies, which after a strong wind phase ($t \le 0.5$ Gyr),
present a low wind phase ($ 0.5 \le t \le 3$ Gyr) and then a long smooth
flux which constitues a hot corona and may be the source of the cooling flow.
In Fig.10 and 11 we plot respectively the behaviour of the diffuse
 gas phase and the
total expelled mass: the presence of three different regimes is evident.

Fig.12 shows the rate of expelled O and Fe, in solar masses per year,
as a function of time: the  lines must not be interpreted
in the sense  of a continuous loss of those elements from the galaxy,
they are only shown for clearness. We note that during the very
first episodes of wind, the amount of expelled oxygen exceeds
iron, due to Type II SNs explosions, while the subsequent
lost gas has a much higher iron to oxygen ratio, for Type Ia SNs.
This must be kept in mind while studying the enrichment of the
intracluster medium: considering the evolution only until
the onset of the first wind means to underestimate the amount of
iron that ellipticals are able to provide to the IGM.

Our model allows a larger number of observational constraints
than others; from the comparison with the data -- so far partial, exaustive
when
the spectrophotometric part of the project will be ready -- we recognize
the importance of phase cycling along the evolution as a key
to understand the present properties of ellipticals, determined, besides
winds, by the gas--clouds--stars conspiracy.

\vfill\eject
\vskip 0.2in
\centerline {\bf 5. DISCUSSION}
\vskip 0.2in

This paper is the first of a series aimed to extend to elliptical galaxies the
self-consistent approach, successful in understanding the evolution of
 spiral galaxies,
and, in particular, their chemical evolution.

The main points of this paper are:

1) the general theoretical picture of ellipticals reveals
systems with high star formation ($\sim 1500 \msun \rm {yr}^{-1}$
compared to the maximum computed SFR in our galaxy of $\sim 10 \msun \rm
 {yr}^{-1}$),
in the initial phases of evolution; this strongly influences both
 dynamics and chemical
enrichment of diffuse matter.

2)  Our model requires high efficiences of gas loss during
the episodes of galactic winds, with typical timescale of $10^4$ years,
to reproduce the observational constraints.

3) The high rates of SNs II in the first evolutionary
phases, consequence of the high SFR  produce
relative chemical abundances in ISM and, consequently,
in the subsequent stellar population, different
from the solar neighbourhood and, in general, spiral galaxies.
This has to be kept in mind in any kind of studies of
ellipticals: the different chemical abundances influence
lifetimes, final stages and spectral energy distribution of stars. Most of the
chemical calibrations derived from observations
in our own galaxy are expected to fail in reproducing
the characteristics of these systems. We will analyse
these effects with more details, particularly concerning
the index $ Mg_{2} $, in a following paper; the detailed
chemical analyses of our model allows to evaluate a new theorical calibration.

4) The stated concept that higher stellar metallicities
imply more recent evolutionary times cannot be applied
to ellipticals; our models present a non-monotic metal
evolution, with episodes of "super metallicities" that
could become observable with expected observational improvements.

5) Studies of stellar evolutionary tracks
with non--solar relative chemical abundances and metal content
are necessary for any study of ellipticals.
The monotic growth of the total metal content with
the $^4$He abundance, supported from observations
and found also by FMPP in their model for the solar neighborhood does not
occur in ellipticals.

6) Within our model we follow the Star Formation Rate
and all the evolution in a self-consistent way
also after the first episode of galactic wind.
In fact, if it is true that the amount of mass
loss from the galaxy is determined mostly from
the first wind or from the winds occurring
during the first Gyr, many other quantities
at the present time are decided from the following
episodes of wind: the  amount of diffuse gas and the consequent SFR
(two of the most important data to compare
with observations) strongly depend on the restoration of
gas from evolving stars and from the galactic winds
occurring after the first Gyr.

\vskip 0.5 cm
{\bf Acknowledgments}\par
We thank F. Matteucci and S.N. Shore for interesting discussions.
We warmly thank Amedeo Tornamb\`e for the help in understanding the present
status of the Type I SNs theory, for his kind calculation  of the
 cumulative number
of SN I tha we used in the model and for his continuous support to
 our researches.
We are grateful to U. Penco for his precious
and competent collaboration to the final development of
the computational code. The comments of the referee, Gillian R. Knapp,
improved the clearness of the paper.

F.F.  acknowledges support from MURST 40\% and MURST 60\%.

\vfill\eject

\nobreak
\vskip 3truecm
\parindent=0pc
\parskip=0pt
\centerline{\bf REFERENCES}
\vskip 2pc


\ref Angeletti, L. \& Giannone, P. 1990, A\&A, 234, 53 (AG)

\ref Angeletti, L. \& Giannone, P. 1991, A\&A, 248, 45

\ref Arimoto, N. 1989 in Evolutionary Phenomena in Galaxies,
ed. J.E.Beckman \& B.E.J. Pagel (Cambridge: C.U.P.), p.323

\ref Arimoto, N. \& Yoshii, Y. 1987, A\&A, 173, 23 (AY)

\ref Arnaud, M., Rothenflug, R., Boulade, O., Vigroux, L. \&
Vangioni-Flam, E. 1992, A\&A, 254, 49



\ref Bertola, F., Burstein, D. \& Buson, L.M. 1992, preprint

\ref Branch, D. \& Nomoto, K. 1986, A\&A, 164, 113

\ref Bregman, J.N., Roberts, M.S. \& Giovannelli, R. 1988,
ApJ, 330, L93

\ref Brocato, E., Matteucci, F., Mazzitelli, I. \& Tornamb\`e, A. 1990,
ApJ, 349, 458

\ref Burkert, A. \& Hensler, G. 1987, in  Nuclear Astrophysics,
 eds. Hillebrandt \sl et al.\rm, Springer-Verlag, p.159


\ref Burstein, D. 1979, ApJ, 232, 74


\ref Canizares, C.R., Fabbiano, G. \& Trinchieri, G. 1987,
ApJ, 312, 503

\ref Cappellaro, E. \& Turatto, M. 1988, A\&A,  190, 10


\ref Ciotti, L., D'Ercole, A., Pellegrini, S. \& Renzini, A. 1991,
ApJ, 376, 380

\ref Cox, D.P. 1972, ApJ, 178, 159

\ref David, L.P., Forman, W. \&  Jones, C. 1990, ApJ, 359, 29

\ref David, L.P., Forman, W. \& Jones, C. 1991a, ApJ, 369, 121

\ref David, L.P., Forman, W. \& Jones, C.:1991b, ApJ,
 380, 39




\ref de Vaucouleurs, G., de Vaucouleurs, A. \& Corwin, H.G. 1976,
Second Reference Catalogue of Bright Galaxies, (Austin: Univ. Texas
Press)

\ref de Vaucouleurs, G., de Vaucouleurs, A., Corwin, H.G.,
Buta, R.J., Paturel, G. \& Fouqu\`e, P. 1991, Third
Reference Catalogue of Bright Galaxies, (New York: Springer
Verlag)

\ref de Young, D.S. 1978, ApJ, 223, 47

\ref Di Serego Alighieri, S., Trinchieri, G. \& Brocato, E. 1990,
in Windows on Galaxies, ed. G.Fabbiano, J.A. Gallagher \&
A. Renzini (Dordrecht: Kluwer), p. 301

\ref Evans, R., van den Bergh, S. \& McClure, R.D. 1989, ApJ, 345, 752

\ref Fabbiano, G. 1989, ARAA, 27, 87

\ref Fabbiano, G., Kim, D.W. \& Trinchieri, G. 1992, ApJ Suppl., 80, 531



\ref Ferrini, F. 1991a, in Evolution of interstellar matter and dynamics of
galaxies, ed. J. Palou\v s (Cambridge U.P.), in press

\ref Ferrini, F. 1991b, in  Chemical and Dynamical evolution
of Galaxies, ed. F. Ferrini, F. Matteucci \& J. Franco (Pisa: ETS), p.511


\ref Ferrini, F. Galli, D. 1988, A\&A, 195, 27

\ref Ferrini, F., Palla, F. \&  Penco, U. 1990, A\&A, 231, 3

\ref Ferrini, F., Matteucci, F., Pardi, C. \& Penco, U. 1992, ApJ,
 387, 138 (FMPP)

\ref Forman, W., Jones, C. \& Tucker, W. 1985, ApJ, 293, 102



\ref Gordon, M.A., 1990, ApJ, 350, L29

\ref Gordon, M.A., 1991, ApJ, 371, 563


\ref Greggio, L., Renzini, A. 1990, ApJ, 364, 35


\ref Haud, U., Joever, M. \& Einasto, J. 1985, IAU Symposium, The Milky
Way Galaxy, eds. H. van Woerden \sl et al. \rm, p.85




\ref Huchtmeier, W.K., Bregman, J.N., Hogg, D.E. \&
Roberts, M.S., 1988, A\&A, 198, L17

\ref Iben, I.Jr. \&  Tutukov, A. 1984, ApJSup, 54, 335

\ref Iben, I.Jr., Nomoto, K., Tornamb\`e, A. \& Tutukov, A. 1987, ApJ,
 317, 717

\ref Jura, M. 1986, ApJ, 306, 483

\ref Jura, M. 1990, in Windows on galaxies, ed.G.Fabbiano, J.A.
Gallagher \& A. Renzini (Dordrecht: Kluwer), p.321

\ref Jura, M., Kim, D., Knapp, G.R. \&
Guhathakurta, P. 1987, ApJ, 312, L11

\ref Kim, D.W., Gutathakurta, P., van Gorkom, J.H., Jura, M. \&
Knapp, G.R., 1988, ApJ, 330, 684




\ref Landau, L. \& Lifshitz, E. 1970, in  Physique theorique --
II:Theorie des champs, ed. MIR, Moscou


\ref Lees, J.F., Knapp, G.R., Rupen, M.P. \&
Phillips, T.G. 1991, ApJ, 379, 177




\ref Matteucci, F. 1992, preprint


\ref Matteucci, F. \& Tornamb\`e, A. 1987, A\&A, 185, 51 (MT)



\ref Nomoto, K., Thielemann, F.K. \& Yokoi, K. 1984, ApJ,
286, 644

\ref Pagel, B.E.J. \& Edmunds, M.G. 1981,  ARAA, 19, 77


\ref Popova E.I., Tutukov, V. \& Yungelson, L.R. 1982,  Astrophys.
 Space Science,  88, 55


\ref Roberts, M.S., Hogg, D.E., Bregman, J.N., Forman, W.R.
\& Jones, C. 1991, ApJ Suppl, 75, 751

\ref Sadler, E.M. 1988, in Cooling Flows in Clusters and
Galaxies, ed. A.C. Fabian (Dordrecht: Kluwer), p. 376

\ref Sadler, E.M. \& Gerhard, O.E. 1985, in New Aspects of
Galaxy Photometry, ed. J.-L. Nieto, (New York: Springer
Verlag), p. 269

\ref Sage, L.J.  \& Galletta, G. 1992, preprint

\ref Saito, M. 1979, Publ. Astron. Soc. Japan, 31, 181


\ref Sandage, A. 1986, A\& A, 161, 89


\ref Schweizer, F. 1987, in
Structure and dynamics of elliptical galaxies,
eds. D. Reidel Publishing Company, Dordrecht, Holland, p.109


\ref Sparks, W.B., Wall, J.V., Thorne, D.J.,
Jorden, P.R. \& van Breda, I.G. et al. 1985,
MNRAS, 217, 87



\ref Thronson, H.A.jr. \& Bally, J. 1987, ApJ, 319, L63

\ref Tinsley, B.M. \& Larson, R.B. 1979, MNRAS, 186, 503



\ref Tornamb\`e, A. 1989, MNRAS, 239, 771

\ref Tully, R.B. 1988, Nearby Galaxies Catalog, ed.
R.B. Tully (Cambridge: Cambridge University Press)




\ref van den Bergh, S. \& Tammann, G.A. 1991, ARAA,
29, 363

\ref Veron-Cetty, M. \& Veron, P. 1988, A\&A, 204, 28

\ref Vigroux, L. 1977, A\&A, 56, 473

\ref Walsh, D.E.P., van Gorkom, J.H., Bies, W.E., Katz, N.,
Knapp, G.R. \& Wallington, S., 1990, ApJ, 352, 532

\ref Whelan, J.C. \&  Iben, I.Jr. 1973, ApJ, 186, 1007

\ref White, R.E.III 1991, ApJ, 367, 69


\ref Yoshii, Y. \& Arimoto, N. 1987, A\&A, 188, 13

\vfill \eject

\centerline{\bf FIGURE CAPTIONS}
\vskip 0.5 cm
Fig. 1 -- Cumulative number of Supernovae Type Ia and Ib
from Tornamb\'e with the IMF adopted in this paper.
\vskip 0.2 cm
Fig. 2 -- Predicted Star Formation Rates ($\msun \,
{\rm yr}^{-1}$) as functions of time both in the internal and external zone.
Time is expressed in units of $10^{9}$ years.
\vskip 0.2 cm
Fig. 3 -- Predicted behaviour as a function of time for the stellar
remnants ($r$), low mass stars ($s$), clouds ($c$) and
diffuse gas ($g$) for the external zone. All the quantities are expressed as
fractions of the total mass.
\vskip 0.2 cm
Fig. 4 -- The same as Fig. 3  but for the internal zone.
\vskip 0.2 cm
Figs. 5 (a,b) -- Predicted age--metallicity relation for $[{\rm Fe/H}]$ and
$[{\rm O/H}]$ (in solar units) in the two zones.
\vskip 0.2 cm
Fig. 6  -- The same as Fig. 5 but for the disk of spiral galaxies.
\vskip 0.2 cm
Fig. 7 -- Predicted age--metallicity relation  in the two zones.
\vskip 0.2 cm
Fig. 8 -- Predicted age--metallicity relation  in spiral galaxies,
both for disk and halo.
\vskip 0.2 cm
Fig. 9 (a,b) -- Predicted $[{\rm O/Fe}]$ vs. $[{\rm Fe/H}]$   relation
for the external region and for the disk of spiral galaxies.
\vskip 0.2 cm
Fig. 10 -- Predicted behaviour as a function of time for the
diffuse gas ($g$) for the two zones.
\vskip 0.2 cm
Fig. 11 -- Predicted behaviour as a function of time for the rate of
expelled mass.
\vskip 0.2 cm
Fig. 12 -- Predicted behaviour as a function of time for the rate
of O and Fe expelled mass.
\vfill\eject


\thicksize=0.8pt
\thinsize=0.4pt

\def\W#1#2{$^{#1}\rm {#2}\ $}


TABLE 1
\medskip
Results of selected models
\bigskip

\begintable

                  | $M_G(0)$    | $M_{dark}$   | $k$  |  $t_{gw}$ | $M_{out}$|
 Fe   | SFR  | $t_{lw}$     | Ia (SNU)  | $g$ |
 $c$    | $s$| $r$  \cr
\W {}A\hfill|     $10^{12}$  | $0$          | $1000$ | 0.35     | 0.68     |
 14.5  |0.2 | 15.79  | 0.05 | 0.002 | 0.002 |0.250 |0.068 \nr
\W {}B\hfill|     $ 10^{12}$| $0$  | $100$ |  0.35     | 0.31     |
 17.4   | 5.1 | 15.54 | 0.08 | 0.011| 0.006 |0.287 | 0.074 \nr
\W {}C\hfill|     $10^{12}$  | $0$   | $1000$ |  0.35   | 0.65     |
 3.4  | 0.6   | 15.79  |  0.005     | 0.002| 0.003 |0.269|   0.077 \nr
\W {}D\hfill|     $2 \cdot 10^{12}$  | $0$| $1000$ |  0.45     |
 0.62     | 17.7          | 0.3  | 15.79  |  0.06     |
0.002| 0.002 | 0.295 | 0.080      \nr
\W {}E\hfill|     $10^{11}$  | $0$
| $1000$ | 0.15  | 0.82 | 0.7 | 0.02  | 16.00 | 0.03 | 0.002 | 0.002 | 0.140
| 0.038 \nr
\W {}F\hfill|     $10^{12}$  | $10^{12}$   | $1000$ |  0.85
 | 0.44    | 21.3 | 1.3  | 15.79  | 0.10   | 0.002| 0.005 |
0.434| 0.116

\endtable


\vfill\eject
\bye